\documentclass[a4paper,11pt]{article}
\usepackage{jheppub}
\usepackage{amssymb}
\usepackage{amsmath}

\usepackage[utf8]{inputenc}
\usepackage[english]{babel}
\usepackage{mathtools}
\usepackage{amsfonts}
\usepackage{amsthm} 
\usepackage{physics} 
\usepackage{comment} 
\usepackage[dvipsnames]{xcolor}
\usepackage{float} 
\usepackage{tikz} 
\usetikzlibrary{shapes,arrows,positioning,automata,backgrounds,calc,er,patterns}
\usepackage[compat=1.0.0]{tikz-feynman}
\usepackage{graphicx}
\usepackage{tabularx}
\usepackage{amssymb}
\usepackage{mathrsfs}
\usepackage{bbm}
\usepackage{accents}
\newcommand{\smalltilde}[1]{%
  \accentset{\raisebox{-0.5ex}{\scalebox{0.58}{$\sim$}}}{#1}%
}

\renewcommand{\d}{\delta}

\renewcommand{\k}{\kappa}

\newcommand{\G}{\Gamma}

\newcommand{\DeclareAutoPairedDelimiter}[3]{%
  \expandafter\DeclarePairedDelimiter\csname Auto\string#1\endcsname{#2}{#3}%
  \begingroup\edef\x{\endgroup
    \noexpand\DeclareRobustCommand{\noexpand#1}{%
      \expandafter\noexpand\csname Auto\string#1\endcsname*}}%
  \x}
\DeclareAutoPairedDelimiter{\p}{(}{)} 

\renewcommand{\tilde}[1]{\widetilde{#1}}

\title{ Spinning Fields in Lorentzian AdS}

\author[a]{David Berenstein}
\author[b]{and Ziyi Li}

\affiliation[a]{Department of Physics, University of California at Santa Barbara, California 93106, USA}
\affiliation[b]{Department of Physics, University of California at Davis, California 95616, USA}

\emailAdd{dberens@physics.ucsb.edu}
\emailAdd{zzyli@ucdavis.edu}

\abstract{We construct the higher spin wave functions in the embedding space of anti-de Sitter Lorentzian spacetime. These wave functions are built from a primary wave functions that has a simple structure expressed in terms of the special conformal generator vector fields in AdS. We compute the eigenvalue of the quadratic Casimir for the symmetric traceless states, and show explicitly that these satisfy the higher spin wave equation. We also demonstrate that these wave functions have the right structure in the flat space limit for massive higher spin fields, and can be used to construct $in$ and $out$ states for scattering processes. Spinning states that become massless in the flat limit are extremely subtle. The problem can be isolated to longitudinal polarizations. }
\newpage

\begin{document} 
\maketitle
\flushbottom

\section{Introduction}
The holographic principle relates quantum gravity in the bulk sapcetime to a non-gravitational system on its boundary \cite{tHooft:1993dmi,Susskind:1994vu}. The most concrete realization of the holographic principle is the AdS/CFT correspondence, which states that quantum gravity/string theory in asymptotically anti-de Sitter spacetime is equivalent to a conformal field theory living on the boundary of the spacetime \cite{Maldacena:1997re}. The observables of the CFT such as the correlation functions can be computed in the bulk gravity theory using the GKPW dictionary \cite{Gubser:1998bc,Witten:1998qj}, and these should be interpreted as computing some type of scattering amplitudes in AdS spacetime \cite{Polchinski:1999ry,Giddings:1999jq,Giddings:1999qu}. However, due to the timelike nature of AdS boundary, these scattering amplitudes are different from the usual notion of $S$ matrix in flat spacetime. In anti-de Sitter spacetime, massive particles cannot reach the boundary at infinity while massless particles bounce back from the boundary periodically. This means that these particles will interact in the bulk continuously and the notion of $in$ and $out$ states no longer applies. However, if the scattering happens at a scale much smaller than that of the AdS radius, the process should essentially become that of the flat spacetime, and one would expect the $S$ matrix of flat space scattering process to emerge in this limit. Indeed, this would be true in any spacetime where one can study quantum fields in curved space time, so long as the region where the process is to take place is much much smaller than any other geometric scale like the curvature.

Although quantum field theory in AdS spacetime essentially reduces to that of the flat spacetime in the $R\to \infty$ limit, where $R$ is the radius of AdS, extrapolation of the flat space scattering amplitude in this limit has proven to be non-trivial. The reason is that in and out states propagate beyond the region where the flat space limit is taken, so the preparation and measurement of these wave packets happen away from the region where the flat space QFT is valid. This is a local-to-global problem, where one has to show that the wave packets that can actually be prepared globally in the AdS setup reduce to wave-packets with the right characteristics in the flat space region, and that contributions to amplitudes from other regions are sufficiently suppressed. In that case the flat space S-matrix is the correct framework to approximate the 
global answer. In a strict double scaling limit, one should recover exactly the flat space S-matrix.

A large body of work has been devoted to studying such a limit, and many subtleties of the limit have been addressed in different formalisms \cite{Gary:2009ae,Gary:2009mi,Fitzpatrick:2011hu,Raju:2012zr,Maldacena:2015iua,Paulos:2016fap,Komatsu:2020sag}. One of the challenging aspects arises from preparation of these states on the boundary, which can be addressed using the HKLL prescription \cite{Hamilton:2006az} to construct local bulk fields \cite{Hijano:2015qja} (see also \cite{Li:2021snj}). However, such approaches involve complicated boundary integrals, and it is difficult to carry out when interactions are introduced. A complementary  approach is to study the  wave function of the bulk fields directly as well as their interactions first, and consider their relationship with boundary preparation later. This is the route taken by \cite{Berenstein:2025tts} for scalar fields, and remarkably, that prescription manifests the flat space limit in a straightforward way especially when treated in the embedding space formalism. The one subtlety is that massless states that are prepared from global AdS with a very similar type of preparation as massive states (there is a subtle scaling of the momentum that one needs to do) end up being solutions with waveforms that have a shape (they are not standard scattering plane waves). In that sense, the flat space limit of this construction does not recover the S-matrix on the nose, but something sufficiently close.

The study of spinning fields in AdS spacetime enjoys a long history \cite{Fradkin:1987ks,Vasiliev:1999ba,Vasiliev:2003ev,Klebanov:2002ja}\footnote{See \cite{Deser:2001pe,Deser:2001us} for the study of partially massless higher spin fields in (A)dS.}, and spinning correlators in the Euclidean language have been studied in \cite{Fernandes:2022con,Costa:2014kfa} and more recently in \cite{Marotta:2024sce}, where it is claimed that the flat limit works well for massive states and that calculations with massless states in AdS also work. Attempts to extract the S-matrix from direct CFT computations date to the original work of Penedones \cite{Penedones:2010ue} and some statements can be derived about how the bulk Feynman rules arise from CFT \cite{Fitzpatrick:2011hu}. There are some subtleties in all these descriptions. First, the flat space limit is supposed to live in 10 or 11 dimensions, not just the flat limit of AdS itself. In that limit, states that start as massive states in the AdS geometry can become massless in the flat limit. So far, most of the calculations in the literature have been done with the exactly massless states in AdS, but we in general expect these other AdS massive states to survive as massless particles. Such states lead to additional complications that have not been considered in detail in the literature. We point out that the problems can be isolated to the longitudinal polarizations of the eventually massless particles, but we are unable to resolve these problems directly without a full theory of the higher dimensions, which we have not  developed yet. 

The main result of the present paper is the construction of the spinning wave functions in AdS spacetime within the same embedding space formalism, aimed at providing the ingredients needed for computing the flat space scattering of higher spin fields along the lines of \cite{Berenstein:2025tts}. These are constructed in the Loretzian theory directly, rather than in Euclidean setups as in \cite{Costa:2014kfa}. The representation theory is slightly different even if they share some characteristics.
One might expect that studying higher spin fields essentially adds an extra layer of difficulty in the flat space limit \cite{Marotta:2024sce}; however, as it turns out, the higher spin wavefunctions possess a remarkably simple representation in the embedding space of AdS, allowing the flat-space limit to be taken in a correspondingly simple manner, at least for massive spinning states. These all become plane wave states in the flat limit expressed in terms of transverse polarizations. 

It can be argued that the Lorentzian embedding space formalism is implicit in the work \cite{Bekaert:2023uve} that treats the problem of constructing Lagrangians for such spinning fields. We instead construct the representation of the conformal group by building the primaries and then taking descendants.
Additional subtleties arise when one considers spinning particles that in the flat space limit become massless, beyond those already present for massless scalars. We will study some of those subtleties but we are not able to resolve them completely.

The paper is organized as follows; In Section \ref{section2}, we review the prescription proposed in \cite{Berenstein:2025tts} and set up the embedding space coordinates. From the representation theory perspective, the flat space limit can be thought simply as the Inonu-Wigner contraction \cite{Inonu:1953sp}, which allows one to focus on the global symmetries of AdS spacetime. Massive primary scalar wave functions in the flat space limit reduce to  plane waves at rest, and to implement scattering processes of moving particles, one needs to boost the wave function, which can be easily implemented in the embedding space of AdS. The boosted wave function reduces to plane waves in the flat space limit, and can be used to construct the flat space $S$ matrix in this limit.  In Section \ref{section3}, we construct the generic form of the higher spin primary wave function in the embedding space, and demonstrate that the special conformal transformation annihilates the wave function. In particular we focus on the symmetric traceless state, and compute the quadratic Casimir of the wave function, thus showing that it indeed belongs to the symmetric traceless representation of $SO(d)$ and solves the wave equation. Along the way, we clarify different definitions of mass used in the literature for symmetric traceless tensors. Finally, we take the flat space limit of the higher spin wave function for the massive spinning particles, and show that they indeed have the same degrees of freedom as the higher spin field in flat space. In Section \ref{section4}, we present explicit formulae for computing the descendants of the higher spin wave function, and compute the first few levels explicitly. We also point out certain subtleties when studying the spinning states of finite $\Delta$ that become massless in the flat space limit. We end with a discussion and some future direction in Section \ref{section5}.

\section{Scalar Fields and their flat space limit}
\label{section2}
Here we review the prescription proposed in \cite{Berenstein:2025tts} for scalar field primaries and also the flat space limit. From the perspective of the representation theory, this emergence of flat space physics from AdS spacetime can be simply viewed as a particular Inonu-Wigner contraction around a local bulk point $p$, say the origin.  Inonu-Wigner contractions also show up in Carrollian theories, which are also built as an avenue to study the flat space limit (see the recent review \cite{Nguyen:2025zhg} and references therein for this literature). These particular relations are beyond the scope of the present paper.

The first step in \cite{Berenstein:2025tts} is to state that single (scalar) particle states in AdS are excitations in a single unitary irreducible representation of the conformal group. Our goal is to show how to build these representations in the bulk.

In conformal field theory, the representations are associated instead to a list of operators inserted at the origin via the operator/state correspondence. Therefore it is customary to express the representation theory in terms of the operator language rather than the state language in the cylinder. Once the origin is chosen, the 
list of operators is constructed from a primary field and its descendants. The primary is ${\cal O}(0)$ and the descendants are associated to the list of operators  $\partial^{[n]}_y {\cal O}(y)|_y=0$.

The primary field is an eigenstate of the dilatation operator $D$ of dimension $\Delta$, and in general it can have rotational quantum numbers in the rotation group that survives by the choice of the origin. These would be the rotations of the $y$ coordinates into each other at $y=0$.
The rotation quantum numbers give a unique  unitary irreducible representation of $SO(d)$ for a conformal primary field theory in $d$ dimensions. This rotation group can  be thought of as the little group of Wigner for these representations. 
Scalar field primaries are singlets under $SO(d)$.
The descendants have conformal dimension $\Delta +k$ where $k>0$ is an integer. In that sense, the primary is the operator of minimal dimension in the representation (a type of lowest weight state).
The condition of being primary is that the special conformal generators acting on ${\cal O}(0)$ vanish, so we can state that $K_i {\cal O}(0)=0$.

The idea now is that we should use the same technique to study irreducible representations of the conformal group in AdS. To each such particle state, we should be able to associate a wave function on AdS that solves the equations of motion of the field in AdS. In that sense, the representation theory is acting on the solutions of the wave equation.

One way to proceed is then to choose one's favorite coordinate system, separate variables and write all the solutions of the wave equations in said coordinates after imposing the correct boundary conditions in the AdS boundary (the fields need to decay in a particular way as in the GKPW dictionary \cite{Gubser:1998bc,Witten:1998qj}).  This is seen for example in \cite{Fitzpatrick:2010zm}. Indeed, the original setup in \cite{Witten:1998qj} used separation of variables in the Poincar\'e slicing of (Euclidean) AdS to perform calculations. 

To adapt the symmetry of the insertion in the operators to the symmetry of the AdS slices, it is usually best to work in global coordinates where the line element looks as follows
\begin{equation}
    ds^2= -\cosh(\rho)^2 dt^2 + d\rho^2 + \sinh(\rho)^2 d\Omega_{d-1}^2
\end{equation}
where now we associate the time $t$ to radial time, the $SO(d)$ symmetry to the sphere coordinates of $\d\Omega_d^2$ and there is the additional radial coordinate $\rho$. The boundary conditions are imposed in the region $\rho\to \infty$.
This coordinate system obscures the other symmetries of AdS.

However, this way of thinking about the problem is not well adapted to the flat space limit, which requires zooming onto a point at $\rho\simeq 0$ at some fixed time $t$, which is located at a coordinate singularity of the coordinate system. Instead, we would like to use flat coordinates near the point $p$ and the Lorentz symmetry of the flat space should be the subgroup of the AdS symmetry group that preserves the chosen point $p$. This includes the symmetries that are not obvious in the global coordinate system above. This gives a different parametrization of the generators of symmetry of AdS in terms of Lorentz generators $L$, and the momenta of flat space ${\mathcal P}$ need to be identified with generators of the symmetry of AdS that move $p$, which we call $\tilde {\mathcal P}$.

The isometry algebra of global AdS$_{d+1}$ consists of the rotation $L$ and translation $\tilde{\mathcal{P}}$ generators, which satisfy the following commutation relations: 
\begin{align}
    \left[L,L\right] \propto L, \quad  \left[L,\tilde{\mathcal{P}}\right] \propto \tilde{\mathcal{P}}, \quad   [\tilde{\mathcal{P}},\tilde{\mathcal{P}}] \propto L
\end{align}
This is almost the same as the local Poincar\'e algebra, except that the translation generators $\tilde{\mathcal{P}}$ do not commute. The idea of the Inonu-Wigner contraction is to introduce a scale parameter $\Lambda$, and rescale the generators to obtain the desired algebra. For the present case, we need: 
\begin{align}
\label{rescalep}
    \mathcal{P}= \frac{\tilde{\mathcal{P}}}{\Lambda}
\end{align}
The commutation relation of the the rescaled generator is then: 
\begin{align}
    \left[\mathcal{P},\mathcal{P}\right] \propto \frac{L}{\Lambda^2}
\end{align}
Taking the $\Lambda \to \infty$ limit while fixing $\mathcal{P}$, we precisely land on the Poincar\'e algebra. Albeit simple, such contractions provide two valuable insights. First, even though the Poincar\'e algebra arises from a particular contraction around a point $p$, the flat space limit can be taken purely from a representation theory perspective, where one can focus on global wave functions in AdS that represent the primary state and take the limit of such states. Secondly, we need to view the eigenvalue of $\tilde{\mathcal{P}}$ in the double-scaling limit as states with fixed momentum $\mathcal{P}$, which are the momentum eigenstates in the flat spacetime. Consequently, one needs to seek AdS space wavefunctions whose flat limit reproduces ordinary plane waves for fields around the point $p$. This is where the embedding space formalism becomes very useful. It permits one to identify $L$ and $\tilde{\mathcal{P}}$ readily, and the conformal group symmetry acts linearly on all coordinates, just like Lorentz transformations do in flat space.

\subsection{Flat Space limit in the Embedding Space Formalism}
The emergence of Poincar\'e algebra from the Inonu-Wigner contraction can be more explicitly seen from the embedding space of AdS, which is naturally represented by a hyperboloid embedding of Lorentzian AdS$_{d+1}$ into   $\mathbb{R}^{d,2}$: 
\begin{align}
    -\p{X^0}^2+\p{X^1}^2+\dotsm -\p{X^{-1}}^2=-1
\end{align}
where we set the radius of AdS spacetime to one. The isometry algebra is simply the 
$SO(d,2)$ Lorentz algebra, whose generator $M_{AB}$ satisfies the following commutation relation: 
\begin{align}
    \left[ M_{AB},M_{CD} \right]=i\p{\eta_{AC}M_{BD}+\eta_{BD}M_{AC}-\eta_{BC}M_{AD}-\eta_{AD}M_{BC}}
\end{align}
where\footnote{Here we are using the convention where the $M_{AB}$ act as self-adjoint operators, hence the factors of $i$.} 
\begin{align}
    M_{AB}=-i\p{X_A\frac{\partial}{\partial X^B}-X_B \frac{\partial}{\partial X^A}}
\end{align}

The Lorentz algebra can then be decomposed into the conformal algebra if we make the following identifications\footnote{See Appendix \ref{Appendixa} for the full algebra.}: 
\begin{align}
\label{decomposecft}
    &M_{0,-1}=-D , \quad  M_{i,0}=\frac{P_i+K_i}{2} \notag \\
    & M_{ij}=J_{ij}, \quad \hspace{4mm} M_{i,-1}=\frac{P_i-K_i}{2i} 
\end{align}
where we have one dilation operator $D$, $d$ translation operators $P_i$, and $d$ special conformal transformations $K_i$ together with $\frac{d(d-1)}{d}$ rotation generators $J_{ij}$. To see how the Poincar\'e algebra arises from the AdS$_{d+1}=$ CFT$_d$ isometry, we note that implementing the scaling (\ref{rescalep}) is the same as the coordinate transformation: 
\begin{align}
\label{coordtrans}
    X^{0,i}=\frac{x^{0,i}}{\Lambda}
\end{align}
where $x^{i,0}$ can be interpreted as the flat spacetime coordinates while keeping $X^{-1}=1$ fixed in the limit, and more precisely, $X^{-1} \simeq 1+O(1/\Lambda^2)\to 1$.
The embedding space generator $M_{-1,i}$ can be seen as the AdS$_{d+1}$ translation generator: 
\begin{align}
  \tilde{\mathcal{P}}_i=  M_{-1,i}=i\p{X^{-1}\partial_{X^i}+X^i \partial_{X^{-1}}}
\end{align}
The locus where we take the flat space limit is restricted to the region around  $p$ ($X^{-1}=1$, $X^0=0,X^i=0$). Taking the limit $X^{-1}=1$, and performing the coordinate transformation (\ref{coordtrans}), we have in the $\Lambda\to \infty$ limit: 
\begin{align}
    \tilde{\mathcal{P}}_i \to \Lambda  \p{i\partial_{x^i}}=\Lambda \mathcal{P}_i
\end{align}
which lands on the correct generator of translation in the flat spacetime. Similarly, the temporal component of the translation generator in AdS can be identified as $\tilde{\mathcal{P}}_0=M_{0,-1}$, where in the flat space limit becomes: 
\begin{align}
    \tilde{\mathcal{P}}_0 \to -\Lambda\p{i \partial_{x^0}}=\Lambda \mathcal{P}_0
\end{align}
Together they form the $d+1$ dimensional translation generator $\mathcal{P}_\mu=i\{ -\partial_{x^0},\partial_{x^i} \}$ in the flat spacetime. The rest of the generator $M_{ij}$ and $M_{i,0}$ are the usual rotational and boost generators of the Poincar\'e algebra which survive in the double scaling limit\footnote{The scale parameter $\Lambda$ cancels for these generators under the coordinate transformation (\ref{coordtrans}).}. 

\subsection{Scalar Wave function}
It is useful to adopt global coordinates for the hyperboloid embedding in the $\mathbb{R}^{d,2}$ spacetime: 
\begin{align}
    X^{0}=\cosh{\rho} \sin{t}, \quad X^{-1}=\cosh{\rho}\cos{t},\quad X^i=\sinh{\rho} \hspace{1mm}n^i
\end{align}
where $n^i$ is the unit vector that parametrizes the a unit $(d-1)$-dimensional sphere. The metric on $\mathbb{R}^{d,2}$ now becomes: 
\begin{align}
    \dd s^2=-\cosh\p{\rho}^2\dd t^2+\dd \rho^2+\sinh\p{\rho}^2\dd \Omega^2_{d-1}
\end{align}
and this is useful to translate problem written on terms of the embedding coordinates into the more familiar global coordinates. Define the holomorphic coordinate: 
\begin{align}
   z=X^{-1}+iX^{0}
\end{align}
and its complex conjugate
\begin{align}
   \bar z=X^{-1}-iX^{0}.
\end{align}
The hyperboloid constraint can then be written as follows
\begin{equation}
    z\bar z-\sum (X^i)^2=1
\end{equation}
The complex variable $z$ satisfies $|z|\geq 1$, so it lives on the complement of a disk. The correct coordinate of 
global AdS is the infinite cover of the disk complement, which is best parametrized by $z=\exp(i \smalltilde{t})$.
The real part of $\smalltilde{t}$ is unconstrained, but the imaginary part is bounded above by zero. That is $\Im  (\smalltilde{t})\leq 0$.
The operators $P,K$ can be seen to be proportional to
\begin{equation}
    K_i\propto z \partial_{X^i}+2 X^i \partial_{\bar z}
\end{equation}
and similarly $P_i \propto K_i^{\dagger}$ is proportional to the complex conjugate vector field.
We can write the general primary scalar wave function as follows: 
\begin{align}
    \phi_{\Delta}=\frac{1}{z^{\Delta}}=\exp(-i \Delta \smalltilde{t})=\frac{1}{\p{\cosh\p{\rho}\exp\p{it}}^{\Delta}}.\label{eq:primary_scalar_function}
\end{align}
The wave function represents a primary state of conformal dimension $\Delta$, and it can be checked immediately that $K_i$ annihilates $\phi_\Delta$ in the embedding space ${\mathbb R}^{2,d}$, where it is actually a harmonic function (it satisfies the equation $\eta^{AB}\partial_{X^A}\partial_{X_B} \phi_{\Delta}=0$). The fact that $K$ acting on $\phi$ is zero guarantees that the wave function is a primary state. 
The right hand side is the restriction of $\phi_{\Delta}$ to the hyperboloid written in global coordinates. 

This wave solution is also obviously rotationally invariant as it does not depend on the $X^{1, \dots, d} $ coordinates.
Unitarity imposes $\Delta>0$, so these solutions have positive energy. In QFT in curved space, these are attached to lowering operators of the quantum field (see for example \cite{Birrell:1982ix}). These are therefore the wave functions of ``in particles" in a scattering process. The complex conjugate representation produces negative energy solutions, which become ``out particles" in a scattering process. We will deal only with the positive energy solutions in this discussion.  

In these holomorphic coordinates for the embedding space we have that 
\begin{equation}
    D= \bar z \partial_{\bar z}- z \partial_z.
\end{equation}

The descendants can be obtained by acting with $P_i$ on the primary. The first few are given (up to normalization) in the table \ref{tab:desc}. Each numerator is a polynomial in the $X^i$ and the combination $z\bar z$. These generalize the idea of harmonic functions being symmetric traceless polynomials.
\begin{table}[ht]
\center
\begin{tabular}{|c|c|}
 \hline  Dimension   & Function  \\
 \hline $\Delta$   & $z^{-\Delta}$ \\
 $\Delta+1$ &   $X^i z^{-\Delta-1}$ \\
$\Delta+2$ & $(X^i X^j-\frac{ z\bar z }{2(\Delta+1)} \delta^i_j )z^{-\Delta-2}$ \\
$\Delta+3$ & $(X^i X^j X^k-\frac{ z\bar z }{2(\Delta+2)} \delta^i_j X^k-\frac{ z\bar z }{2(\Delta+2)} \delta^i_k X^i -\frac{ z\bar z }{2(\Delta+2)} \delta^k_j X^i )z^{-\Delta-3}$
\\
 \hline
\end{tabular}
\caption{First few descendants of the primary $\phi_\Delta$ }\label{tab:desc}
\end{table}
These can be expressed directly in the $\rho, t, n^i$ coordinates if we want to, but the expressions are rather cumbersome.

In the semiclassical limit where $\Delta$ is large, the wave function is essentially localized around the center $\rho=0$ of AdS and can be thought of as a particle at rest trapped in the AdS potential well.
When we take the  flat limit we get that 
\begin{equation}
\psi_\Delta = \frac 1{(1+i x^0/\Lambda)^{\Delta}} \to \exp(-i (\Delta/\Lambda) x^0) 
\end{equation}
and we notice that so long as $\Delta/\Lambda= m$ is held fixed, we end up with a finite energy state in the flat space limit, with energy equal to $m$. This wave function is at rest (the space momentum vanishes) and it is clearly a plane wave.

The descendant states are generated by acting with $P_i$ on the primary state. However, in the flat space limit, $P_i\sim \Lambda \partial_{x^i}$. This means that $P_i$ will need to be rescaled in the flat limit, i.e. $P_i/\Lambda$. Thus, the notion of the descendants does not produce new states in this limit.
This is also another reason we need to focus on the global wave function of the primary state. Thus, in order to build the correct quantum field representation in the flat limit, one needs to consider the boosted wave functions that represent moving particles \cite{Berenstein:2025tts}. This is thinking a la Wigner to build unitary irreducible representations of the Poincar\'e group, rather than in the usual representation theory of the Conformal group language. Obviously, the primary wave function survives as a particle at rest and it also has the minimal energy of all single particle states with the same mass. 

The boost can be easily implemented for the embedding space coordinate $z$:
\begin{align}
    z\to \xi =X^{-1}+i\p{\cosh\p{\eta}X^0+\sinh\p{\eta}m_iX^i}
\end{align}
where we have boosted $z$ with rapidity $\eta$ and direction $m^i$ (this is a unit vector). The wave function can be written covariantly by introducing an embedding space complex null vector $\kappa^A$: 
\begin{align}
    \kappa_A=\p{1,-ik_\mu}, \quad k_\mu=\p{-\cosh\p{\eta},-\sinh\p{\eta}m^i}
\end{align}
where the boosted wave function is now: 
\begin{align}
    \phi_{\Delta,\k}=\frac{1}{\xi^\Delta}\equiv \frac{1}{\p{\k \cdot X}^{\Delta}}
\end{align}
The boosted wavefunctions with different $\kappa$ are non-orthogonal. For massive states, however, their overlaps vanish in the flat-space limit, so they form an orthogonal basis. More explicitly, these boosted wave function become plane waves in the double scaling limit: 
\begin{align}
    \phi_{\Delta,\k}=\p{1-ik_\mu x^\mu/\Lambda}^{-\Delta}\sim \exp\p{i\Delta k_\mu x^\mu/\Lambda}
\end{align}
In the double scaling limit, $\Delta=\Lambda m_{\text{flat}}$, and we recognize $p_{\mu}=m_{\text{flat}}k_\mu$ as the momentum for the plane wave. Notice these have positive energy as their angular frequency is  $\omega= -m_{flat}k_0 >0$.

The out states are simply the complex conjugate of the boosted wave function above. Thus, the scattering process can be set up in the embedding space, where the integral is over the entire space $\mathbb{R}^{d,2}$, and the hyerboliod constraint can be added as a delta-function. Indeed, the integral is essentially over the plane waves in the flat space limit and we recover the usual delta function of momentum conservation times the flat space S-matrix \cite{Berenstein:2025tts}. 

Moreover, the preparation of these states on the boundary does not require the complicated boundary integral prescription (HKLL construction) as shown explicitly in \cite{Berenstein:2025tts}. This is essentially due to the fact that these boosted wave functions are primary states with respect to some boosted special conformal transformation $\tilde{K}_i$ (See section \ref{masslesspin}), which annihilates them. The existence of such $\tilde{K}_i$ indicates that the operator/state is a primary with respect to an insertion localized on the boundary at a single (complex) time and position.

When considering states with $\Delta$ finite, in the flat space limit these go to massless particles. As shown in \cite{Berenstein:2025tts}, when we boost the primary states enough in a double scaling sense, we get a non-trivial set of solutions of the massless equations of motion that have a shape. The primaries themselves correspond to a notion of massless particles at rest, which does not make sense from unitary representation theory of the Lorentz group for massless particles. We will revisit this construction later when we discuss the flat limit of spinning particles at finite $\Delta$.

We can ask however, is there a sense in which the primary and the descendants have a meaning in the flat space limit? The answer is actually yes, but we have to get away from the massless particle picture.

For that we need to think what happens when we consider the wave function in the double scaling limit. We obtain a wave packet with no time dependence $\phi\simeq z^{-\Delta} \to 1$. It can be interpreted as a shift of the expectation value of the field $\phi$ in the region we want to study. So long as we double scale the amplitude of the background field so that the shift survives in the double scaling limit, we get a flat space field theory with a different vacuum. For example, we can have the dilaton in string theory have a different value near the flat space region than on the boundary.

Similarly, we can keep the first descendant so that $\phi \simeq X^i$ becomes a background field configuration with a non-constant profile. The background is still a solution of the massless field equations and grows towards infinity, so in flat space it would have infinite energy.
When we boost this solution, we can also get a time dependent profile and by superposition, and we can eliminate the spatial gradient if we want to. At the next order, we would get for example a {\em quadrupole} harmonic profile in $X$, like one has for the scalar potential in electromagnetism inside a cavity.
Basically, the solutions can survive as zero modes of the background field which we can turn on perturbatively and study susceptibilities of the field theory in flat space to the presence of those background fields.

\section{The Higher Spin Wave Functions}
\label{section3}
The main result of the present paper is the construction of higher spin primary wave functions. The others follow immediately by taking descendants (acting with vector fields on the primary wave function). 
From the representation theory perspective, these wave functions are primary states in the CFT, which have the minimum energy and are annihilated by the special conformal transformations $K_i$. Solving these wave functions directly in AdS$_{d+1}$ can be rather cumbersome even for the Proca field (massive spin-1 field), where the d'Alambertian mixes different components of the vector field and extra care is required in decoupling the radial differential equation \cite{Fernandes:2022con,Lopes:2024ofy}. Here we take a different route in dealing with that problem. Instead of solving the radial differential equation directly in AdS$_{d+1}$, we construct the global wave function for higher spin fields directly in the embedding space, and perform several consistency checks for such wave function, which in turn demonstrates that it solves the wave equation. Formally, wave functions can be handled by the formalism of \cite{Bekaert:2023uve} with formal power series and a formulation to get to a Lagrangian. However, stating the problem with a primary and its descendants and showing that it solves the correct  PDE is all we need. This is more analogous to the construction of fields from irreducible representations of the Lorentz group as in Weinberg \cite{Weinberg:1995mt}\footnote{See especially the discussion of how the wave equations are produced for massive spin one particles directly from the representation theory in chapter 5.}.

We claim that such a primary wave function takes the form: 
\begin{align}
    \mathcal{O}_{i_1i_2\cdots i_\ell}=\phi K_{i_1}\otimes K_{i_2} \cdots \otimes K_{i_\ell}
\end{align}
where $\phi=z^{-\Delta}$ is simply the scalar wave function and the $K$s are simply the vector fields associated to the special conformal transformations in $AdS$. One important point is that the $K$ are vector fields in AdS itself (they are parallel to the hyperboloid manifold). Hence, one does not need to check for vanishing of transversality conditions, they are automatic.
That way the spin of the wave function gets transformed into a specific tensor field of AdS.
It is simple to see that such wave function represents a primary state:
\begin{align}
    K \mathcal{O}=\p{K \phi}K\otimes\cdots\otimes K+\phi\big([K,K \otimes\cdots\otimes K]\big)=0
\end{align}
where we have used the fact that $K$ annihilates the scalar wave function, and it commutes with all the other $K$, in the sense of Lie derivatives. These are thus primary states with energy $\Delta'=\Delta-\ell$, since each $K$ lowers the conformal weight by one unit. Having shown that these wave functions are indeed primary, we need to check that the state belongs to the correct representation of little group $SO(d)$. For that, we need to decompose the tensor products into rotation group multiplets.

\subsection{Symmetric Traceless Tensor}
Let's focus on the symmetric traceless wave function and work out the eigenvalues of the Casimir of such states.
What is important about the Casimir is that as an operator it has two derivatives. Hence it is a natural {\em wave equation operator} and directly shows that the fields will satisfy second order differential equations and can therefore lead to a notion of mass (similar to how fields in flat space have a mass and a spin).

The computation of the Casimir will demonstrate that they indeed represent the appropriate symmetric traceless states. Once the primaries are constructed, all the descendants will automatically satisfy the same second order differential equation as they will share the Casimir. That basically builds the correct fields in the treatment of Weinberg described above.
This amounts to computing the eigenvalue for the quadratic Casimir operator: 
\begin{align}
\label{secondcasmir}
    \frac{1}{2}M_{AB}M^{AB}=D(D-d)-P_iK_i+\frac{1}{2}J_{ij}J_{ij}
\end{align}
where the eigenvalue $\lambda$ defined as: 
\begin{align}
    \frac{1}{2}M_{AB}M_{AB}\hspace{1mm} \mathcal{O}_{i_1i_2\cdots i_\ell}=\lambda \hspace{1mm} \mathcal{O}_{i_1i_2\cdots i_\ell}
\end{align}
Using the commutation relation of the conformal algebra as well as $\left[D,\phi\right]=\Delta \phi,\left[K_i,\phi\right]=0$, the first term in (\ref{secondcasmir}) gives:  
\begin{align}
    D(D-d)\mathcal{O}_{1 \cdots \ell}=\Delta'(\Delta'-d)
\end{align}
where $\Delta'=\Delta-\ell$, while the second term annihilates the state. The last term is more cumbersome to work out, but follows straightforwardly from the conformal algebra and the traceless condition ${\cal O}_{i_1i_1i_3, \dots i_\ell}=0$. We provide detailed computation in Appendix \ref{Appendixa}, where it is shown that: 
\begin{align}
    \frac{1}{2}J_{mn}J_{mn} \hspace{1mm} \mathcal{O}_{i_1i_2\cdots i_\ell}=\ell\p{\ell+d-2} \mathcal{O}_{i_1i_2\cdots i_\ell}
\end{align}
Indeed, this is the correct eigenvalue for states belonging in the symmetric traceless spin $\ell$ representation of $SO(d)$, whose quadratic Casimir is: 
\begin{align}
    \lambda=\Delta'(\Delta'-d)+\ell\p{\ell+d-2}
\end{align}
The mass of the higher spin field however, requires some care, since in AdS spacetime, the mass is an ambiguous concept due to coupling to the curvature. For a scalar field, the quadratic Casimir can be written in terms of local AdS$_{d+1}$ coordinates as: 
\begin{align}
     \frac{1}{2}M_{AB}M^{AB}\hspace{1mm}\phi=\nabla^2_{\text{AdS}}\phi
\end{align}
The wave equation $\nabla^2_{\text{AdS}}\phi=m^2\phi$ then simply yields the mass relation:
\begin{align}
\label{scalarmass}
    \Delta\p{\Delta-d}=m^2
\end{align}
This however, is no longer true for higher spin fields, and we cannot simply equate the eigenvalue of the quadratic Casimir with the mass. There are two extra contribution that change the definition of the mass. The first comes from a constant shift in the quadratic Casimir acting on higher spin-$\ell$ field \cite{Pilch:1984xx}: 
\begin{align}
\label{shift1}
    \frac{1}{2}M_{AB}M^{AB}\hspace{1mm} \mathcal{O}_{i_1i_2\cdots i_\ell}=\big[\nabla^2_{\text{AdS}}+\ell\p{\ell+d-1}\big] \mathcal{O}_{i_1i_2\cdots i_\ell}
\end{align}
Replacing the left hand side of the above equation with the eigenvalue $\lambda$, and define the mass as the eigenvalue of the operator $\nabla^2_{\text{AdS}}$, we have\footnote{This definition of mass was used in \cite{Hijano:2015zsa, Hijano:2015qja}.}: 
\begin{align}
    m^2= \Delta'\p{\Delta'-d}-\ell 
\end{align}
However, this definition of mass does not take into account of the interaction between the tensor fields with the curvature, since the wave equation satisfied by the tensor fields in AdS$_{d+1}$ also contains a constant shift from the mass due to fact that covariant derivatives do not commute when acting on non-scalar fields. It can be shown that the wave equation for a generic symmetric traceless tensor is given by \cite{Giombi:2013yva}:
\begin{align}
\label{eom2}
    \left[\nabla_{\text{AdS}}^2+2-\p{\ell-2}\p{\ell+d-3}\right] \mathcal{O}_{i_1i_2\cdots i_\ell}=m^2 \mathcal{O}_{i_1i_2\cdots i_\ell}
\end{align}
Using the eigenvalue $\lambda$ together with (\ref{shift1}), the above definition of mass gives: 
\begin{align}
\label{mass}
    \p{\Delta'+\ell-2}\p{\Delta'-\ell-d+2}=m^2
\end{align}
which is the standard definition used in the AdS/CFT contexts. For example, if $m^2=0$ and $\ell=1$ ( a massless spin one particle), we get $(\Delta'-1)(\Delta'-1-d+2)=0$
so that we recover the usual property that a spin one massless field is dual to operators of dimension $d-1$
(we need to choose the second root $\Delta'=d-1$), which is a conserved current. The other solution of $\Delta'$ for massless particles, with $\Delta'=1$ is non-normalizable and usually corresponds to  source term
in the GKPW dictionary.
Similarly, a massless spin two particle requires $\Delta'=d-2+2=d$ which is the usual dimension for the dual stress tensor.  Next, we need to show that our wavefunction does not generate any new states when traces are being taken, and we shall illustrate this with the spin-2 field.

\subsection{The Spin-2 Field and Traces}
\label{traces}
In this subsection, we will compute explicitly how the spin-2 field $\mathcal{O}_{ \ell m}=\phi K_{(\ell} \otimes K_{m)}$ transforms under the rotation generator $\frac{1}{2}J_{ij}J_{ij}$, and we show that requiring the trace to be nonzero, i.e. $\eta^{\ell m} K_{\ell} \otimes K_{m}\neq 0 $, we simply land back on a scalar field again. We wish to compute:
\begin{align*}
   \frac{1}{2} \left[J_{ij}, \left[J_{ij},K_{\ell}\otimes K_{m} \right]    \right]
\end{align*}
Since we have a tracesless spin-two tensor field, where $K_iK_i=0$\footnote{As usual, the repeated index is summed over.}. The inner commutator is then: 
\begin{align*}
   \left[J_{ij},K_{\ell}\otimes K_{m} \right]&= \left[ J_{ij},K_{\ell }\right]\otimes K_{m}+K_{\ell}\otimes\left[J_{ij},K_{m} \right] \\
   &=\underbrace{i\p{\delta_{i\ell}K_j-\delta_{j\ell}K_i}\otimes K_m}_{(1)}+\underbrace{iK_\ell\otimes \p{\delta_{im}K_j-\delta_{jm}K_i}}_{(2)}
\end{align*}
The total commutator is then: 
\begin{align*}
    \left[J_{ij}, \left[J_{ij},K_{\ell}\otimes K_{m} \right]    \right]&=\left[ J_{ij},(1)    \right]+\left[J_{ij},(2)\right] 
\end{align*}
where we have 
\begin{align}
\label{spin2commutation}
    \left[ J_{ij},(1)    \right]&=\bigg(\p{2(d-1)K_{\ell}\otimes K_{m}}+(2K_{m}\otimes K_{\ell}-2\delta_{\ell m}K_i\otimes K_i)\bigg) \notag \\
    \left[J_{ij},(2)\right] &=\bigg(\p{2(d-1)K_{\ell}\otimes K_{m}}+(2K_{m}\otimes K_{\ell}-2\delta_{\ell m}K_j\otimes K_j)\bigg)
\end{align}
Since $K_iK_i=0$, and we have symmetrized over the indices, the final result is given by: 
\begin{align}
    \frac{1}{2} \left[J_{ij}, \left[J_{ij},K_{\ell}\otimes K_{m} \right]    \right]=2d
\end{align}
This is indeed the correct eigenvalue of $\frac{1}{2}J_{ij}J_{ji}$ for $\ell=2$. We would also like to take the trace of above wave function: $\phi \delta^{m \ell}K_mK_\ell$. More importantly, we need to show that the trace does not transform under rotation. We can simply take the trace in equation (\ref{spin2commutation}). Using the fact that $\delta^{\ell m}\delta_{\ell m}=d$, we see that the right-hand sides of both equations are zero, and it is a simple statement that the operator does not transform under the rotation, and thus behaving like a scalar with eigenvalue $\lambda=\p{\Delta-2}\p{\Delta-d-2}$, that is, a scalar primary of dimension $\Delta'=\Delta -2$.

This can be seen to follow straightforwardly from the fact that the metric tensor acting on the vector fields $K$ is given by $g(K_i,K_j)= z^2 \delta_{ij}$, so the trace can be taken with the metric tensor of the manifold, which is an invariant of the conformal symmetry and we get a wave function without the $K$ insertions exactly like the one we discussed in equation \eqref{eq:primary_scalar_function}. This can be done generically to get rid of excess $K$ vector fields when we can take a trace.
In the flat limit such scalar wave functions would look like a tensor with components $\exp(ik_\mu x^\mu) (\eta^{\mu\nu} +k^\mu k^\nu/m^2)$ which is built from polynomials of $k_\mu$ and the metric like we would expect in Feynman rules. This can be seen as
$(\eta^{\mu\nu}+ 1/m^2\partial_\mu\partial_\nu)\phi(x)$, which is a local operator built from $\phi$ and its derivatives.

\subsection{Flat Space limit of the Higher Spin Wave Function}
As discussed in section \ref{section2}, in order to build the correct quantum field representation in the flat limit, one needs to consider the boosted wave functions. Having shown that the scalar part of the wave function essentially becomes plane waves, we need to show that the $K$s also reduce to the appropriate vector in the flat limit. Under the double scaling limit, this can be trivially worked out: 
\begin{align}
    K_i=-\Lambda \partial_{x_i}-ix^0\partial_{x_i}-ix_i\partial_{x^0} \sim -\Lambda \partial^i
\end{align}
Similar to $P_i$, we also need to rescale $K_i$, which gives in the flat space limit $K_i\to -\partial_{x^i}$. This vector has $d$ independent components and indeed satisfies the right degree of freedom for the polarization vector of a massive vector field in $d+1$ dimension. These can be used to construct the \textit{in} states for the scattering processes.
The answer for the primary is given by
\begin{equation}
 \mathcal{O}_{i_1i_2\cdots i_\ell}=\phi K_{i_1}\otimes K_{i_2} \cdots \otimes K_{i_\ell}
 \to \exp(-i \Delta/\Lambda x^0) \partial_{x^{i_1}} \otimes \dots \partial_{x^{i_\ell}}
\end{equation}
which is a massive particle at rest with polarization amplitude in the directions indicated by the $i_1i_2\cdots i_\ell$ (here for illustration we are ignoring the symmetry properties of the indices).
The wave vector $k_\mu = (-\omega, 0)$ shows that the amplitudes are transverse to the time direction $k_\mu A^{\mu\dots }=0$.
The transformation properties of the representation of the $i_1i_2\cdots i_\ell$ are exactly the spin degrees of freedom in the Wigner construction of states at rest and are obviously in an irreducible of the $SO(d)$ little group if the primary wave function was in such an irreducible representation.

For $out$ states, similar to the scalar wave functions, should be constructed by the Hermitian conjugate of the $in$ state, which is given by: 
\begin{align}
    \mathcal{O}_{out}=\p{\phi_{\Delta,\k}K_{i_1}\otimes \cdots \otimes K_{i_\ell}}^{\dagger}=\phi^*_{\Delta,\kappa^*} P_{i_1}\otimes \cdots \otimes P_{i_\ell}
\end{align}
These wave functions are also (dual) primary states, which now are annihilated by $K_i{}^\dagger=P_i$\footnote{Note that $\phi^*$ is annihilated by $P_i$ and $P$s commute among themselves.}. Moreover, these states belong to the complex conjugate representation as the $in$ wave function, since the commutation relation between $P$ and $J$ is identical to that of $K$ and $J$, which means that they will have the same quadratic Casimir of the little group $SO(d)$ and the conformal group itself. Note that the notion of Hermitian conjugate survives in the flat space limit, since $K_i{}^\dagger=(-\Lambda\partial_{x^i})^\dagger=\Lambda\partial_{x^i}=P_i$, where we have used the fact that the position derivative is anti-Hermitian. 

Boosting the massive states is straightforward as we can do the boost in flat space directly (in the $x^i$ coordinates after using $X^{-1}=1$) and we get the correct polarizations for boosted particles. 

In the spirit of the discussion in \cite{Berenstein:2025tts}, it is obvious that the S-matrix will be the right answer for scattering problems in the limit, as the localization properties of the wave packets near the origin are essentially identical for massive spinning  particles at large $\Delta$. They are controlled by the profile of $z^{-\Delta}
\simeq 1/\cosh(\rho)^\Delta$.

\section{Descendants of Higher Spin Wave Function}
\label{section4}
The descendants of the symmetric traceless state can be obtained by acting $P_i$ on the primary state, and carry energy $E=\Delta'+N$, where $N$ represents the number of time $P_i$ acts on the state. The Verma module for the symmetric traceless primary state can be expressed as: $\mathcal{V}_{\Delta',\ell}=\oplus_{N}\mathcal{V}_{\Delta',\ell,N}$, where
\begin{align}
    \mathcal{V}_{\Delta',\ell,N}=\text{Span}\{P_{i_1}\cdots P_{i_N}|\mathcal{O}_{j_1j_2\cdots j_\ell}\rangle\}
\end{align}
The indices are all symmetrized and schematically, this can be written as: 
\begin{align}
    P^N \p{\phi \mathbb{K}}=\sum_{k=0}^{N}\frac{\G\p{N+1}}{\G\p{k+1}\Gamma\p{N-k+1}}\p{P^{n-k}\phi} \cdot \p{[P^k, \mathbb{K}]}
\end{align}
where we have defined $\mathbb{K}_{i_1\cdots i_\ell}=K_{i_1}\otimes\cdots \otimes K_{i_\ell}$. These can be made explicit if necessary and replace the description of the problem as done by separation of variables. It is implicit in the {\em spin} quantum numbers of the descendants.

The action of $P$ on the scalar part of the wave function is simply $P\phi$, and $P$ acting on $\mathbb{K}$ is then given by: 
\begin{align}
    P_j \mathbb{K}_{i_1\cdots i_\ell}=[P_j,\mathbb{K}_{i_1\cdots i_\ell}]=\sum_{m=1}^{\ell}K_{i_1}\cdots K_{i_{m-1}}[P_j,K_{i_m}]K_{i_{m+1}}\cdots K_{i_\ell}
\end{align}
 Moreover, we can write compactly $P^k \mathbb{K}$ as: 
 \begin{align}
     [P^k,\mathbb{K}]=\sum_{s=1}^{k} \frac{\p{-1}^{s-1}\G\p{k+1}}{\G\p{s+1}\Gamma\p{k-s+1}}P^{k-s}\underbrace{\bigg[ \big[P,\cdots [P,\mathbb{K}]\big]\bigg]}_{s-\text{nested commutators}}
 \end{align}
Combining these two equations, we have: 
\begin{align}
    P^N \p{\phi \mathbb{K}}=\sum_{k=0}^{N}\sum_{s=1}^{k} \frac{\p{-1}^{s-1}\G\p{N+1}\p{P^{n-k}\phi}}{\G\p{s+1}\Gamma\p{N-k+1}\Gamma\p{k-s+1}} P^{k-s}\underbrace{\bigg[ \big[P,\cdots [P,\mathbb{K}]\big]\bigg]}_{s-\text{nested commutators}}
\end{align}
Note that for these descendants, the most general ones can be written by taking traces of pairs of $P$s, given by: 
\begin{align}
    P_{i_1} \cdots P_{i_N}\p{P^2}^n \mathcal{O}_{j_1j_2\cdots j_\ell}
\end{align}
and the energy of such state is then simply $E=\Delta'+N+2n$. For a Proca field, the normal modes are simply given by: 
\begin{align}
    \omega=E=N+2n+\frac{1}{2}\p{d+\sqrt{\p{d-2}^2+4m^2}}
\end{align}
where we solved $\Delta'$ using (\ref{mass}) with $\ell=1$, which is the same as the results obtained in \cite{Lopes:2024ofy} by directly solving the radial differential equation.

\subsection{Descendants and Massless Higher Spin Fields }
Consider the wave function of a spin-1 field given by $z^{-\Delta}K_i$. A massless vector field in the bulk corresponds to a conserved current on the boundary given by $\partial_iJ^i=0$. From the embedding space formalism, such conservation corresponds to computing $P_i\p{z^{-\Delta}K_i}$: 
\begin{align}
    P_i\p{z^{-\Delta}K_i}=-2z^{-\Delta}\p{\Delta r_i K_i+d\cdot D}
\end{align}
where we have defined $r_i=X_i/z$. Such state is a descendant of the spin-1 vector state. To see if such state is a primary, we compute: 
\begin{align}
    K_j\left[P_i\p{z^{-\Delta}K_i}\right]=-2z^{-\Delta}\p{-\Delta K_j+dK_j}\label{eq:prim_desc}
\end{align}
where we have used $K_i(r_j)=-\d_{ij}$. We see that the primary state condition corresponds to: 
\begin{align}
    \Delta=d \implies \Delta'=d-1 
\end{align}
where we have used $\Delta'=\Delta-1$. This is precisely the conformal dimension for a conserved current in the CFT$_d$. One can similarly verify this with higher spin field fields. A massless graviton in the bulk corresponds to a symmetric traceless conserved stress energy tensor $\partial_iT^{ij}=0$. The wave function of a general symmetric traceless state is given by:
\begin{align}
\label{symmetricspin2}
    \mathcal{O}_{ij}=\frac{z^{-\Delta}}{2}\p{K_i\otimes K_j+K_j\otimes K_i-\frac{2}{d}K_m\otimes K_m \d_{ij}}
\end{align}
The first (null) descendant is given by: $P_i\mathcal{O}_{ij}$, and a straightforward calculation shows that the null state condition is given by (See Appendix \ref{appendeixc}): 
\begin{align}
    K_{\ell}\p{P_i \mathcal{O}_{ij}}=0 \implies \Delta'=d
\end{align}
It is more convenient if we use the following set of coordinate transformations:
\begin{align}
\label{newcoordinate}
    X^i=r^i e^{i\smalltilde{t}}, \hspace{0.5cm}\quad z=e^{i\smalltilde{t}}, \hspace{0.5cm}\quad \bar{z}=z^{-1}+z\p{ r^ir_i}
\end{align}
where now the rotation generator $J$ can be expressed as: 
\begin{align}
    J_{ij}=i\p{r_iK_j-r_jK_i}, \quad K_i=-\partial_{r^i}
\end{align}
We can thus show that the first descendant with $\Delta'=d$ can be expressed more compactly as: 
\begin{align}
P_i\p{\mathcal{O}_{ij}}=-\frac{(d+2)(d-1)}{d\cdot z^{\Delta}}\p{\tilde \Delta\otimes K_j+K_j\otimes \tilde \Delta}
\end{align}
where we have defined the vector field 
\begin{equation}
    \tilde \Delta = D {+}r^i K^i
\end{equation}
It is easy to show that $[\tilde \Delta, K_i]=0$. There is a theorem on manifolds that states that given a manifold of dimension $d+1$, one can have at most $d+1$ linearly independent commuting fields and they are usually associated to a coordinate system $\partial_{w^\mu}$. In our case, these coordinates are taken to be $r^i, \smalltilde{t}= \log(z)/i$, where the transformations from the embedding space coordinates are given explicitly in (\ref{newcoordinate}). It is easy to show that $K_i \hspace{0.5mm} \smalltilde{t}=0$, and we can thus write:
\begin{eqnarray}
    K_i \equiv -\partial_{r^i} \\
    \tilde \Delta\equiv i\partial_{\smalltilde{t}} 
\end{eqnarray}
In these coordinates, the metric is given by
\begin{equation}
    ds^2= -\dd \smalltilde{t}\hspace{1mm}^2+ \exp\p{2i \smalltilde{t}} \dd r^i \dd r^i
\end{equation}
which looks very similar to the Poincar\'e slicing of Euclidean AdS if we analytically continue the coordinates so that $i \smalltilde{t}= \tau$ is a real variable. Alternatively, this looks like the flat slicing of de Sitter space cosmology if we ignore the factor of $i$ in the exponential in  the metric for the $r^i$ coordinates. 

In the work \cite{Marotta:2024sce} on the flat space limit of Euclidean correlators, the analytic continuation is implicit in the discussion of how to turn Euclidean answers to Lorentzian answers. Now we will describe some issues of the analytic continuation that are not obvious.

First, neither of $r^i, \smalltilde{t}$ is real, but they are useful for manipulations.
Indeed, when we check for domains of validity we find that $|r^i|<1$ and that $\arg(r^i) = -\Re\p{\smalltilde{t}}$.
The first inequality $|r^i|<1$ is not obvious from the naive analytic continuation. In the work \cite{Marotta:2024sce}, all calculations proceed first by writing plane waves in the directions of $r^i$ which produces Bessel Kernels. These are not orthogonal if we impose the constraint $|r^i|<1$. Also, $\tilde t$ is not purely imaginary and the flat limit is in the 
edge of the domain of definition of the $z$ variable (the edge of the disk). We believe this is tied to the fact that the Poincar\'e slicing in AdS only covers a patch of the geometry so the analytic continuation that defines the flat space limit is exotic. In any case, the continuation from Euclidean to Lorentzian  signature is not trivial. Although this doesn't seem to affect the exact flat space limit because in a sense $r^i$ becomes infinitesimal, it should definitely affect the leading deviation away from the limit.

The primary wave functions of the tensor field we have discussed can be written in this coordinate system  as
\begin{equation}
    \exp(- i \Delta \smalltilde{t}) \partial_{r^{i_1}}\otimes \cdots \otimes \partial_{r^{i_\ell}}
\end{equation}
and the condition of being a primary reduces to $\partial_{r^i} \phi=0$, which is really simple. In the flat limit this is what becomes $\partial_i \Phi=0$ that declares that the massive spinning particle is at rest.

We now want to specialize to vector fields.
The vector field $\tilde \Delta$, or equivalently $\partial_{\smalltilde{t}}$ is also a primary field.
This shows up when we take differentials. For example, for a primary scalar field we have that
\begin{equation}
\dd z^{-\Delta} \propto z^{-\Delta} \dd \smalltilde{t}
\end{equation}
and when we lower the indices we get that we have a polarization proportional to 
\begin{equation}
    z^{-\Delta} \tilde \Delta 
\end{equation}
When we considered earlier if a descendant of a vector field was also primary in \eqref{eq:prim_desc}, we found a vector field exactly as above. This arises from gauge transformations where we take vector fields (usually represented as forms) and switch the gauge to $A\to A+d\phi(z)$. That is, the descendants that are also primary fields arise from gauge transformations acting on primary wave functions of scalars. These disappear when we consider the field strength instead
\begin{equation}
F_i\propto z^{-\Delta}\left( \partial_{\smalltilde{t}}\otimes \partial_{r_i} -\partial_{r_i}\otimes  \partial_{\smalltilde{t}}\right)
\end{equation}
which transforms under the conformal group in the same way as the primary with only one derivative (it has the same conformal weight and rotation group quantum numbers) . The descendant $P^i F_i$ in this setup actually vanishes
identically. 

We can also check the following. The primary wave function for differential 1-forms (vectors) is given by for example:
\begin{equation}
    A= z^{-\Delta+2} dr^1 
\end{equation}
and the field strength is
\begin{equation}
    dA = i (-\Delta+2) z^{-\Delta+1} d\tilde t \wedge (z dr^1)
\end{equation}
The dual field strength is 
\begin{equation}
    \ ^* dA = i (-\Delta+2) z^{-\Delta+1}  z dr^2 \wedge \dots \wedge z dr^d 
\end{equation}
where we are using normalized vielbeins $e^0 = d\tilde t, e^i= z d r^i$. That is
\begin{equation}
    \ ^* dA = i (-\Delta+2) z^{-\Delta+d}   dr^2 \wedge \dots \wedge  dr^d 
\end{equation}
Exactly when $\Delta=d$, we find that $d^ *(dA)=0$ we satisfy the equations of motion of a massless vector field. That is exactly the condition that makes the dimension of the current $[J]=d-1$, so that we have  a conservation law on the boundary (there is a null state). 
Notice that we seem to have now two options for massless form field. Those where the primary is electric given by $A\simeq z^{-\Delta+2k} d r^1 \wedge \dots d r^k$, so that the field strength looks like an electric field in these coordinates, namely $ d\tilde t\wedge d r^1 \wedge \dots d r^k$, and we can also consider a magnetic primary, where $F \simeq  d r^1 \wedge \dots d r^k\wedge d r^{k+1}$, where the field strength looks like a magnetic field (it has no $d\tilde t$ component). This second solution is of dimension $k+1$. 

For a spin one particle ( a vector field), the primary of that form  would correspond to a potential field given by $A\simeq r^i d r^j$ which is technically not annihilated by $K_i$, only the field strength is. This looks like a descendant of the other boundary conditions for spin one, which also has $m^2=0$. Indeed, since that representation is not expected to be unitary, there is no contradiction of a descendant  being primary and the descendant state producing something physical (a background field). It just should not be interpreted as a one particle state. This bears some resemblance to the problem of tachyons in quantum field theory, when we expand around an unstable saddle. An exponentially growing field is not a one particle state, but it is perfectly acceptable as a dynamical background field. The corresponding lowest weight primary of the representation would be $A=dr^i$ which looks like a gauge transformation of the trivial configuration.

\subsection{The flat space limit of massless spinning particles}
\label{masslesspin}

We can now try to understand how to take the flat space limit of a massless field. 
We start with the observation in \cite{Berenstein:2025tts} that we need to do a large boost so that 
the energy of the particle state is finite in the double scaling limit.
We take it along the direction $X^1$ for simplicity.

This proceeds by looking at the boosted notion of $z$
\begin{equation}
    \xi = X^{-1}+i (\cosh \eta X^0+\sinh \eta X^1)
\end{equation}
which is equal to 
\begin{equation}
    \xi = X^{-1}+\frac{i}{2} (\exp \eta/\Lambda  x^0+\exp \eta x^1/\Lambda) +O(\exp(-\eta))
\end{equation}
The double scaling limit requires $\exp(\eta)/\Lambda$ to stay finite. 
Thus in the doubles scaling limit we have that 
\begin{equation}
    \xi = 1+ i n_\mu x^\mu
\end{equation}
where $n^\mu$ is a null vector in the flat space limit. The primary wavefunction of a scalar becomes
\begin{equation}
\psi_\Delta \to \frac 1{\xi^\Delta}= \frac 1{(1+i n_\mu x^\mu)^\Delta}
\end{equation}
It can be seen that $\psi_\Delta$ satisfies the equations of motion of a massless field exactly because $n_\mu$ is null. These are the states that survive for a scalar.

For a vector (tensor), we need to decompose the directions along $x^1$ and $x_\perp$ systematically (basically the direction indicated by $n_\mu$ and the orthogonal spatial complement that does not involve $X^0$). That is, we decompose the polarizations into transverse polarization and longitudinal polarization.

How do we do this? Let us start with the transverse polarizations.
We should begin with 
\begin{equation}
    K_{i,\perp}\simeq z\partial_{X^i} +2 X^i \partial_{\bar z}
\end{equation}
now we pass to the coordinates $x_\perp$ and we notice that the two terms scale differently with $\Lambda$ and only the first contributes.
For example for vector particles we start with
\begin{equation}
    K_i \to \xi \partial_{x^i}+O(1/\Lambda^2)
\end{equation}
and the primary wave functions go to
\begin{equation}
\phi K_i \to \xi^{-\Delta+1} \partial_{x^i}
\end{equation}
where we replace $\xi= 1+ i n_\mu x^\mu$ in the expressions. In that sense, this works identically to the scalars.
Again, one can check that all these solve the equations of motion of a massless vector particle because $n_i=0$ when $i\perp n$, the direction of the wave profile. The orthogonality condition is expressed in flat space as $\epsilon_\mu n^\mu=0$.  
This generalizes to higher spin straightforwardly.

Now we need to worry about the longitudinal polarizations. For that case, it is better to look at $K_1$ as follows:
\begin{equation}
    K_1= X^{-1}\partial_{X^1}+X^1 \partial_{X^{-1}} +i X^0 \partial_{X^1}+i X^1\partial_{X^0}
\end{equation}
What is important is that under a boost in the $01$ directions, the generator $i X^0 \partial_{X^1}+i X^1\partial_{X^0}$ is invariant, that is, it doesn't scale with factors of $]\exp(\eta)$. Similarly, $X^{-1}$ is invariant and we can ignore the terms that have $\partial_{X^{-1}}$ in the flat limit. Therefore we just need to look at the term that actually  boosts, so that 
\begin{equation}
\tilde K=  X^{-1} \cosh \eta \partial_{X^1}-X^{-1} \sinh(\eta)  \partial_{X^0} \to \frac 12 X^{-1}  \exp(\eta)
(\partial_{X_1}-\partial_{X^0})
\end{equation}
we get then that
\begin{equation}
\phi K_1 \to \xi^{-\Delta} \exp(\eta) \partial_{X^+}    
\end{equation}
which is indeed longitudinally polarized. 
Notice that when we do the usual rescaling of $X^+= x^+/\Lambda$, the longitudinal polarization ends up scaling differently than the transverse polarizations. 
For massless AdS particles this does not matter as the longitudinal polarization is pure gauge and decouples.
However, for particles with finite $\Delta$ that are not massless, the extra factor of $\exp(\eta)$ seems to indicate that there are kinematic divergences in the flat limit that need to be handled with care. In these cases we do expect an additional polarization, which is the goldstone mode that would be eaten up to produce a finite mass in the AdS units. The equivalence theorem of particle physics (see for example \cite{Peskin:1995ev}, sec 21.2 and also the discussion in \cite{Donoghue:1992dd}) then tells us that the longitudinal polarization should be replaced by the corresponding goldstone boson. Since the symmetry breaking is of a scale $\Lambda^{-1}$, one would hope that the scattering amplitudes involving this extra factor of $1/\Lambda$ conspire to give finite answers in the limit. In a sense, the equivalence theorem is indicating that we should undo the Higgsing and study the vector boson as a massless field with the goldstone mode treated independently and the mixing between them suppressed by $1/\Lambda$. Then  it should exactly disappear in the limit.
Studying this in detail is beyond the scope of the present paper. 
We believe that the gauge issues indicated in \cite{Marotta:2024sce} correspond exactly to this phenomenon.

What we have indicated with this analysis is that in the flat limit the longitudinal polarizations need to be tackled with extreme care, and to do it correctly is not obvious at all. Since in most cases of AdS/CFT, the flat limit does not only involve the AdS directions but also the sphere (or additional compact directions beyond AdS), the flat limit needs to be handled in higher dimensions to get it right. It seems unlikely that dealing with AdS alone is enough. 

Consider now flat limits as described in the work  \cite{Komatsu:2020sag}. There they show that propagators also go to the flat limit correctly. Also van Rees has a program to replace the LSZ formula by studying QFT on AdS \cite{vanRees:2022zmr}. The results of our work show that, adapted to that program, we can also deal with massless fields, and we are not constrained to study theories with a gap. In such setups, one only expects a finite number of massless spin one particles at most as the boundary theory does not have a stress tensor. The issues we discussed above simply do not apply. However, understanding the decoupling of the unphysical modes can give rise to unphysical singularities in intermediate steps, exactly because the scaling of the longitudinal polarization is different.

\section{Conclusion}
\label{section5}
In this work, we constructed explicitly the wave function for higher spin fields in AdS$_{d+1}$. These wave functions process a simple structure in the embedding space $\mathbb{R}^{d,2}$ in terms of the special conformal generators. We have shown explicitly that for the symmetric traceless case, these wave functions belong to the symmetric traceless representation of $SO(d)$, by computing its quadratic Casimir, and in turn demonstrating that they satisfy the correct wave equation with the appropriate definition of mass. We have further shown that for the massive case in the flat space limit, these wave functions give rise to the correct notion of higher spin fields in flat spacetime, and shown how they can be used to construct $in$ and $out$ states for flat space along the lines of  \cite{Berenstein:2025tts}. These furnish the correct in/out states for the program  of van Rees \cite{vanRees:2022zmr}. We are able to also write massless states in this setup, and just as for scalars, the massless states end up having a non-trivial shape.
These are very similar to the celestial amplitude waveforms (see the reviews \cite{Pasterski:2021rjz,Raclariu:2021zjz,Pasterski:2021raf} and references therein).
This connection needs to be studied in more detail.

Because we found the exact solutions for the primaries and the limit waveforms, our methods substantially reduce the difficulties of solving for the Kaluza-Klein modes of spinning fields in AdS that arise from separation of variables methods. Our methods as written only work on AdS because the symmetry generators are not only used to impose constraints, but they also show explicitly in the waveforms.

We have also found that states with spin that are light in AdS but become massless in the flat limit (they have finite dimension $\Delta$) are very hard to treat correctly. The problem is localized exclusively on the longitudinal polarizations. Transverse polarizations have no issues. We believe that the correct treatment for these states is to undo the Higgs mechanism that gave them mass and to keep track of the symmetry by including the goldstone modes separately as massless scalars. This is suggested by the equivalence theorem in particle physics. These issues arise exactly in AdS/CFT when we consider the the flat limit should end up in a higher dimensional flat space (this issue is also raised in \cite{Fontanella:2025tbs}). Since we have not included a theory of what to do with the sphere $S^5$ in ${\cal N}=4$ SYM for example, we can not tackle 
this problem yet, but is currently under consideration.

Additionally, we have not tackled spinors, which are also fundamental for particle physics. In a  conformal field theory it is tempting to replace the special conformal generators by special supersymmetry generators $K\to S$ and write spinning  primaries as $z^{-\Delta} S$. The one issue is that $S$ is a differential operator acting on superspace, so it is not obvious that this is producing a waveform in the same sense that $K$ as a vector field does.
However, if the superprimary is a bosonic field, supersymmetry actions on the primary waveform should produce the correct wave functions for the spinor descendants and one should then be able to forget the supersymmetry to tackle fermions as wavefunctions on their own. We are looking at this possibility.

Finally, we have not looked carefully at the extrapolate dictionary for all of these states and what that tells us about the signatures of the S-matrix physics in the dual theory. This is currently under investigation.

\acknowledgments
D.B. would like to thank N. Craig,  S. Giddings, G. Horowitz, D. Marolf, J. Sim\'on, M. Srednicki for very useful discussions. The work of D.B. was
supported in part by the Department of Energy under grant DE-SC 0011702.

\appendix
\section{Computation of the quadratic Casimir}
\label{Appendixa}
In this appendix, we compute the quadratic Casimir acting on the symmetric traceless higher spin wave function $\mathcal{O}_{12\cdots \ell}$\footnote{Here we use numbers $1,2,\cdots$ as a replacement for the tensor index.}. It is useful to write out the conformal algebra obtained from the decomposition (\ref{decomposecft}): 
\begin{align}
\label{conformal algebra}
    &\left[ D,P_i \right]=P_i \notag\\
    &\left[ D,K_i \right]=-K_i \notag\\
    &\left[ J_{ij},P_k \right]=i\p{\delta_{ik}P_j-\delta_{jk}P_i} \notag \\
    &\left[ J_{ij},K_k \right]=i\p{\delta_{ik}K_j -\delta_{jk}K_i}\notag \\
    &\left[K_i,P_j \right]=-2\p{iJ_{ij}-\delta_{ij}D}\notag \\
    &\left[J_{ij},J_{k\ell} \right]=i\p{\delta_{ik}J_{j\ell}+\delta_{j\ell}J_{ik}-\delta_{jk}J_{i\ell}-\delta_{i\ell}J_{jk}}
\end{align}
The quadratic Casimir discussed above can be similarly computed as: 
\begin{align}
\label{qc}
    \frac{1}{2}M_{AB}M^{AB}=&M_{0,-1}M^{0,-1}+M_{i0}M^{i0}+M_{i,-1}M^{i,-1}+\frac{1}{2}J_{ij}J_{ij}  \notag \\
    =&M_{0,-1}M_{0,-1}-M_{i,0}M_{i,0}-M_{i,-1}M_{i,-1}+\frac{1}{2}J_{ij}J^{ij} \notag \\
    =&D^2-\frac{1}{2}P_iK_i-\frac{1}{2}K_iP_i+\frac{1}{2}J_{ij}J_{ij} \notag \\
    =&D\p{D-d}-P_iK_i+\frac{1}{2}J_{ij}J_{ij}
\end{align}
Note that for a spin $\ell$ field the first term just gives us: 
\begin{align}
    D(D-d)\mathcal{O}_{1 \cdots \ell}=(\Delta-\ell)(\Delta-\ell-d)
\end{align}
and all we need to show is that it has the right eigenvalue under the rotation generator $\frac{1}{2}J_{ij}J_{ij}$\footnote{In general, the wave function is given by the symmetrized tensor product of $K$ minus the trace, e.g. (\ref{symmetricspin2}). However, as shown explicitly in section \ref{traces}, the trace will not transform under rotation and we will omit them here (or simply use $K_iK_i=0$). }: 
\begin{align}
     \frac{1}{2} \left[J_{ij}, \left[J_{ij}, \phi K_{1}\otimes \cdots \otimes K_{\ell}  \right]    \right]=  \frac{1}{2}\phi \left[J_{ij}, \left[J_{ij}, K_{1}\otimes \cdots \otimes K_{\ell}  \right]    \right]
\end{align}
The inner commutator gives: 
\begin{align}
    &\left[J_{ij}, K_{1}\otimes \cdots \otimes K_{\ell}  \right]= \notag \\ 
   & =\left[J_{ij},K_1 \right]\otimes K_2\otimes \dots \otimes K_{\ell}+\left[J_{ij},K_2 \right]\otimes K_{1}\otimes K_3 \otimes \cdots \otimes K_{\ell}+\cdots +\left[ J_{ij},K_{\ell}  \right] \otimes K_{1}\otimes \cdots \otimes K_{\ell-1} \notag
\end{align}
Note that there are in total $\ell$ such terms. Each term will contribute: 
\begin{align}
\label{hh}
    \left[J_{ij}, \left[J_{ij},K_1 \right]\otimes K_2\otimes \dots \otimes K_{\ell}  \right]&=\left[J_{ij}, \left[J_{ij},K_1 \right] \right] \otimes K_2\otimes \dots \otimes K_{\ell}+\left[J_{ij},K_1 \right] \otimes \left[J_{ij}, K_2\otimes \dots \otimes K_{\ell}  \right] \notag \\
    &=2\p{d-1}K_{1}\otimes \cdots \otimes K_{\ell}+2\p{\ell-1}K_1 \otimes \cdots \otimes K_{\ell}
\end{align}
where the first term simply follows from the fact that: 
\begin{align}
    \left[J_{ij}, \left[J_{ij},K_1 \right] \right] =2\p{d-1} K_1 
\end{align}
The second term requires some work. Let us first illustrate this with spin $\ell=3$: 
\begin{align}
    \left[J_{ij},K_1 \right] \left[J_{ij}, K_2\otimes K_{3}  \right]&=i\p{\delta_{i1}K_j-\delta_{j1}K_i}\otimes i \left[ \p{\delta_{i2}K_j-\delta_{j2}K_i}K_3+K_2\p{\delta_{i3}K_j-\delta_{j3}K_i} \right] \notag \\
    &=2K_1\otimes K_2\otimes K_3+2K_1\otimes K_2 \otimes K_3
\end{align}
It is clear that the first term $\p{\delta_{i1}K_j-\delta_{j1}K_i}$ tensored with $\left[J,K\right]$ will contribute a factor of two: $2\cdot K_1\otimes \cdots \otimes K_\ell$, and there are $\ell-1$ $\left[J,K\right]$ terms left, where in this case $\ell-1=2$. Thus, in case of general higher spin $\ell$ fields, we have: 
\begin{align}
    \left[J_{ij},K_1 \right] \otimes \left[J_{ij}, K_2\otimes \dots \otimes K_{\ell}  \right]=2\p{\ell-1}K_1 \otimes \cdots \otimes K_{\ell} 
\end{align}        
Thus, multiplying (\ref{hh}) by $\ell$ terms gives us: 
\begin{align}
    \frac{1}{2} \left[J_{ij}, \left[J_{ij}, K_{1}\otimes \cdots \otimes K_{\ell}  \right]    \right]=&\frac{1}{2}\ell\big(2\p{d-1}+2\p{\ell-1} \big)K_{1}\otimes \cdots \otimes K_{\ell} \notag \\
    &=\ell\p{\ell+d-2}K_{1}\otimes \cdots \otimes K_{\ell} 
\end{align}
The eigenvalue for the quadratic Casimir for our wave function with symmetric traceless spin $\ell$ is then: 
\begin{align}
\label{eigenvalues}
    \frac{1}{2}M_{AB}M_{BA}\mathcal{O}_{1 \cdots \ell}=\p{ \Delta'\p{\Delta'-d}+\ell\p{\ell+d-2}}\mathcal{O}_{1 \cdots \ell}
\end{align}
which is indeed the expected value we need for a symmetric traceless tensor state of spin $\ell$ in the $SO(d)$.

\section{Spin-2 Field}
\label{appendeixc}
In this appendix, we elaborate some computations for the spin-2 field: 
\begin{align}
    \mathcal{O}_{ij}=\frac{z^{-\Delta}}{2}\p{K_i\otimes K_j+K_j\otimes K_i-\frac{2}{d}K_m\otimes K_m \d_{ij}}
\end{align}
The massless spin-2 field in the bulk is dual to a conserved stress energy tensor on the boundary, and the conservation condition corresponds to computing the first descendant $P_i\p{\mathcal{O}_{ij}}$, which we do term by term: 
\begin{align}
\label{spin2descendant}
    P_i\p{z^{-\Delta}K_i\otimes K_j}&=\p{P_iz^{-\Delta}}K_i\otimes K_j+z^{-\Delta}[P_i,K_i]\otimes K_j+z^{-\Delta}K_i\otimes [P_i,K_j]  \notag\\
    &=2 z^{-\Delta} \p{-\Delta r_i K_i\otimes K_j-d D\otimes K_j+K_i\otimes iJ_{ji}-K_j\otimes D } \notag \\
     P_i\p{z^{-\Delta}K_j\otimes K_i}&=\p{P_iz^{-\Delta}}K_j\otimes K_i+z^{-\Delta}[P_i,K_j]\otimes K_i+z^{-\Delta}K_j\otimes [P_i,K_i] \notag \\
     &=2z^{-\Delta}\p{-\Delta K_j\otimes r_iK_i-dK_j\otimes D+iJ_{ji}\otimes K_i-D\otimes K_j}  \notag \\
     \frac{2}{d}P_j\p{z^{-\Delta}K_m\otimes K_m}&= 2z^{-\Delta}\p{-\frac{2\Delta}{d}r_j K_m\otimes K_m+ \frac{2}{d}\p{iJ_{mj}\otimes K_m+K_m\otimes iJ_{mj}}-\frac{2}{d}\p{D\otimes K_j+K_j\otimes D}}
\end{align}
In order to satisfy the null state condition, such a state has to be a primary state which is annihilated by $K_\ell$, which gives: 
\begin{align*}
    K_\ell\left[P_i\p{\mathcal{O}_{ij}}\right]=z^{-\Delta}\p{(\Delta-d-2)\p{K_j \otimes K_\ell+K_\ell \otimes K_j}+\p{2-\frac{2\Delta}{d}+\frac{4}{d}}\delta_{j\ell}K_m\otimes K_m}
\end{align*}
We see that this will vanish only if $\Delta=d+2$. Using the coordinates (\ref{newcoordinate}), and expressing the rotational generator in terms of the special conformal transformation, we can express (\ref{spin2descendant}) with $\Delta=d+2$ more compactly as: 
\begin{align}
P_i\p{\mathcal{O}_{ij}}=-\frac{(d+2)(d-1)}{d\cdot z^{\Delta}}\p{\tilde \Delta\otimes K_j+K_j\otimes \tilde \Delta}
\end{align}
where we have $\tilde{\Delta}=D+r^iK^i$.

\bibliographystyle{jhep}

\bibliography{refs}

\end{document}